\documentclass[]{spie}  %>>> use for US letter paper
\usepackage{amsmath,amsfonts,amssymb}
\usepackage{graphicx}
\usepackage[colorlinks=true, allcolors=blue]{hyperref}

\title{The Black Hole Explorer (BHEX):\\Preliminary Antenna Design}

\author[a]{~~~T.K. Sridharan}
\author[a]{R. Lehmensiek}
\author[b]{D. Marrone}
\author[c]{P. Cheimets}
\author[c]{M. Freeman}
\author[d,e,f]{P. Galison}
\author[c]{J. Houston}
\author[c]{M. Johnson}
\author[g]{M. Silver}

\affil[a]{National Radio Astronomy Observatory, Charlottesville, VA 22903, USA }
\affil[b]{University of Arizona, Steward Observatory, Tucson, AZ 85719, USA}
\affil[c]{Center for Astrophysics $|$ Harvard \& Smithsonian, Cambridge, MA 02138, USA}
\affil[d]{Black Hole Initiative at Harvard University, 20 Garden Street, Cambridge, MA 02138, USA}
\affil[e]{Department of History of Science, Harvard University, Cambridge, MA 02138, USA}
\affil[f]{Department of Physics, Harvard University, Cambridge, MA 02138, USA}
\affil[g]{MIT Lincoln Laboratory, Lexington, MA 02421}

\authorinfo{Further author information: (Send correspondence to T.K.S)\\T.K.S: E-mail: tksridha@nrao.edu}

\begin{document} 
\maketitle

\begin{abstract}
We present the basic design of a large, light weight, spaceborne antenna for the Black Hole Explorer (BHEX)  space Very Long Baseline Interferometry (space-VLBI) mission, achieving high efficiency operation at mm/sub-mm wavelengths. An introductory overview of the mission and its science background are provided. The BHEX mission targets fundamental black hole physics and astrophysics enabled by the detection of the finely structured image feature around black holes known as the photon ring, theoretically expected due to light orbiting the black hole before reaching the observer. Interferometer baselines much longer than an earth diameter are necessary to attain the spatial resolution required to detect the photon ring, leading to a space component. The science goals require  high sensitivity observations at mm/sub-mm wavelengths, placing stringent constraints on antenna performance. The design approach described, seeks to balance the antenna aperture, volume and mass constraints of the NASA Explorers mission opportunity profile and the desired high performance. A 3.5 m aperture with a 40 $\mu$m surface rms is targeted. Currently, a symmetric, dual reflector, axially displaced ellipse (Gregorian ring focus) optical design and metallized carbon fiber reinforced plastic (CFRP) sandwich construction have been chosen to deliver high efficiency and light weight. Further exploration of design choices and parameter space and reflector shaping studies are in progress. 
\end{abstract}


\section{Science Background and Mission Overview}
\label{sec:intro}  % \label{} allows reference to this section

Black holes are of profound importance both in astrophysics and fundamental physics. Supermassive black holes (SMBH) exist at the centers of most galaxies. They connect phenomena over many orders of magnitude in spatial scales: the fundamental spacetime singularity within the event horizon on solar system size scales, to the millions of light years scales of the whole galaxy, through powerful feedback delivered by relativistic jets. Study of simple, but hard to measure, fundamental black hole properties - masses and spins - and physics on event horizon scales are key to understanding the above processes. BHEX aims to meet these challenges.

The strong gravity near black holes can bend light paths to the point of photons going into orbit. Light orbiting the black hole before arriving at the observer results in a finely structured observable image feature: a nested set of rings called the photon ring \cite{Johnson_2020}.  The photon ring consists of sub-rings indexed by the number of half-orbits executed, $n$. Observations of the photon ring provide enable BHEX to measure and study black hole spins, jets and mass distribution.

While hard to image, the event horizon scale photon ring structure has a strong characteristic ringing interferometric signature detectable (even) by a single interferometer baseline of variable length. The high spatial resolution (few $\mu$arcsec), or equivalently the long interferometer baselines, required to detect this signature unequivocally ( $n \ge 1$), can only be realized with space-VLBI. Ground observations by the Event Horizon Telescope (EHT) have detected the  $n = 0$ emission.\cite{EHTC_M87_I, EHTC_SgrA_I}  Spinning (Kerr) black holes produce non-circular sub-rings and a corresponding signature on orthogonal baselines, providing a spin measurement path. 

To fulfill these goals, BHEX is designed to achieve the highest spatial resolution in history of 6-9 $\mu$arcsec to detect the photon ring, obtain direct measurements of a black hole masses and spins with M87 \& Sgr A$^*$ as primary targets, and to reveal the $n=0$ shadows of dozens of SMBHs. BHEX incorporates high speed laser data downlink at $\sim$100 Gb/s and simultaneous dual band operation in the 80-106.6 GHz \& 240-320 GHz frequency bands. An inclined orbit of at least $\sim$25000 km and a non-maser frequency/timing reference are part of the baseline design targeting a NASA Explorers class mission. Ground based observations, limited to $\sim$345 GHz, cannot attain the required spatial resolutions. At lower $\sim$22 GHz frequencies (e.g. RadioAstron), synchrotron self-absorption obscures the photon ring features. The selected 240-320 GHz band allows access to event horizon scale spatial resolutions and unobscured near-black-hole emission. The 80-106 GHz band enables frequency phase transfer for interferometer station delay tracking and correction. These requirements place stringent constraints on the antenna that collects the electromagnetic field signals at the space location. Deeper discussions of the science and mission concepts, requirements and technology assessments\cite{BHEX_Johnson_2024, BHEX_Marrone_2024, BHEX_Peretz_2024} and subsystem concepts can be found elsewhere in these proceedings.
%BHEX_Galison_2024,BHEX_Wang_2024,BHEX_Rana_2024,BHEX_Srinivasan_2024,BHEX_Tomio_2024,BHEX_Issaoun_2024}

\section{Antenna Technologies}

While the antenna subsystem constitutes the largest (and the first) component of the science payload, its mass must be kept as low as possible. The combination of the large size ($\sim$ 3.5 m) and the required frequency range of 80 - 320 GHz ( up to $<$ 1 mm wavelength) falls in a gap in the large spaceborne antenna technology landscape. Light weight antennas with unfurled/deployable metal mesh fabric surfaces are typical in the X band, e.g. communication and synthetic aperture radar (SAR) applications. Although the mesh technology is feasible in Ka band and offered up to V band, carbon fiber reinforced plastic (CFRP) sandwich reflectors are more common for high gain antennas (HGA) at higher frequencies, e.g., the Europa Clipper and the Roman Space Telescope communication HGAs, and the Planck telescope and the EarthCARE CPR (Cloud Profiling Radar) main reflectors. Heavy mirrors are necessary in the THz and Far Infrared (FIR) bands to achieve large apertures with the required high surface precision, e.g., Herschel. We carried out a survey of available technologies to identify the path appropriate for the BHEX antenna which operates in the intervening mm/sub-mm wavelength range, with a large aperture. Representative examples of available technologies and realizations are shown in the table (not exhaustive), which combines available information from the literature, vendors and our estimates.

\begin{table}[ht]
\caption{Space Antenna Technologies} 
\label{tab:AntennaTechnologies}
\small
\begin{center} 
\begin{tabular}{|l|l|l|l|l|l|l|l|l|} 
\hline
\rule[-1ex]{0pt}{3.5ex}Model/ & Vendor & Size & Surface& Area& Efficiency$^4$& Mass& Bands & Technology \\
  Mission &  &  & accuracy& density& $\eta_{RuzeGHz}$& (3.5 m) &  &\\
          &  & m & $\mu$m & kg/m$^{-2}$& 86/230/320& kg&  &\\
\hline
\rule[-1ex]{0pt}{3.5ex}Herschel & Astrium- & 3.5 & $<$ 10&25 &1 &210 &THz/FIR &metallized SiC\\
&Airbus& & & & & & &\\
\hline
\rule[-1ex]{0pt}{3.5ex}Planck & Astrium- & 1.6$\times$&7.5-50 &13 &$>$ 0.98$^5$ & 125 &mm - THz &metallized\\
&Airbus&1.9 & & & & & &CFRP sandwich\\
\hline
\rule[-1ex]{0pt}{3.5ex}BHEX & - & 3.5 & $<$ 40& $\sim$ 5 &0.98/0.85/0.75 &$\sim$ 50 &mm/sub-mm & metallized\\
(target) & & & & & & & & CFRP sandwich\\
\hline
EarthCARE & NEC & 2.5 & $<$ 60 & - & 0.95/-/- & - & mm & CFRP sandwich \\
\hline
\rule[-1ex]{0pt}{3.5ex}Europa & AASC$^1$ & 3 & $<$ 150&4 &0.75/-/- &38 &Ka & CFRP sandwich\\
Clipper& & & & & & & &\\
\hline
\rule[-1ex]{0pt}{3.5ex}FMR & L3 Harris  & 3.2 &$<$ 270 &2.6 &0.38/-/- &$\sim$ 25 &upto V & mesh\\
\hline
\rule[-1ex]{0pt}{3.5ex}PTR$^{2,3}$ & L3 Harris & 2-22&$<$ 250 & 2&- &$\sim$ 20 &Ka &mesh\\
\hline
\rule[-1ex]{0pt}{3.5ex}SDR$^3$ & Airbus & 5 &$<$ 250 &2.2 &- &$\sim$ 20 &X &CFRP\\
\hline 
\end{tabular}
\end{center}
\tiny{
$^1$Advanced Aerospace Structures Corp. $^2$Perimeter Truss Reflector $^3$deployed $^4$estimates excluding reflectivity $^5$based on predicted whole surface rms}
\end{table}

Light weight antenna realization without performance compromise - the largest aperture and the highest surface precision - is essential to fit within the NASA Explorers class mission profile envelope. The BHEX approach is to extend metallized CFRP sandwich technology to the required higher frequencies and larger sizes through industry partnership. Thermoelastic deformations are expected to be kept within limits through the use of a sunshield to avoid direct sunshine on the antenna. A passive wavefront correcting custom subreflector, tailored to the realized primary reflector to ease its large scale manufacturing surface error requirements and active thermal control to limit thermoelastic deformations are also under consideration.  The industry capabilities, expected development timelines and costs were assessed through a Request for Information (RFI) issued in late 2023, leading to the identification of the feasibility,  technology and target parameters for the BHEX antenna which are included in the table.  

\section{Antenna Optical Design}

The primary optical design consideration is high efficiency. For a VLBI instrument, where the final science beam is computationally constructed on the ground by correlating recorded signals from individual stations, the station primary beam pattern is not important and the peak gains decide the sensitivity. Accordingly, as an on-axis design maximizes the projected effective area for a given physical area, offset designs such as the one employed in Planck were discarded. Axially symmetric antennas are also easier to accommodate with launchers from a mechanical stand point and with downstream optics and recievers\cite{BHEX_Tong_2024}. Non-circular apertures were also considered to leverage the aspect ratio of payload fairings to accommodate a larger aperture, but discarded as their aperture efficiency is considerably lower than an axially symmetric system.   

\begin{figure} [ht]
   \begin{center}
   \begin{tabular}{c} %% tabular useful for creating an array of images 
   \includegraphics[height=6cm]{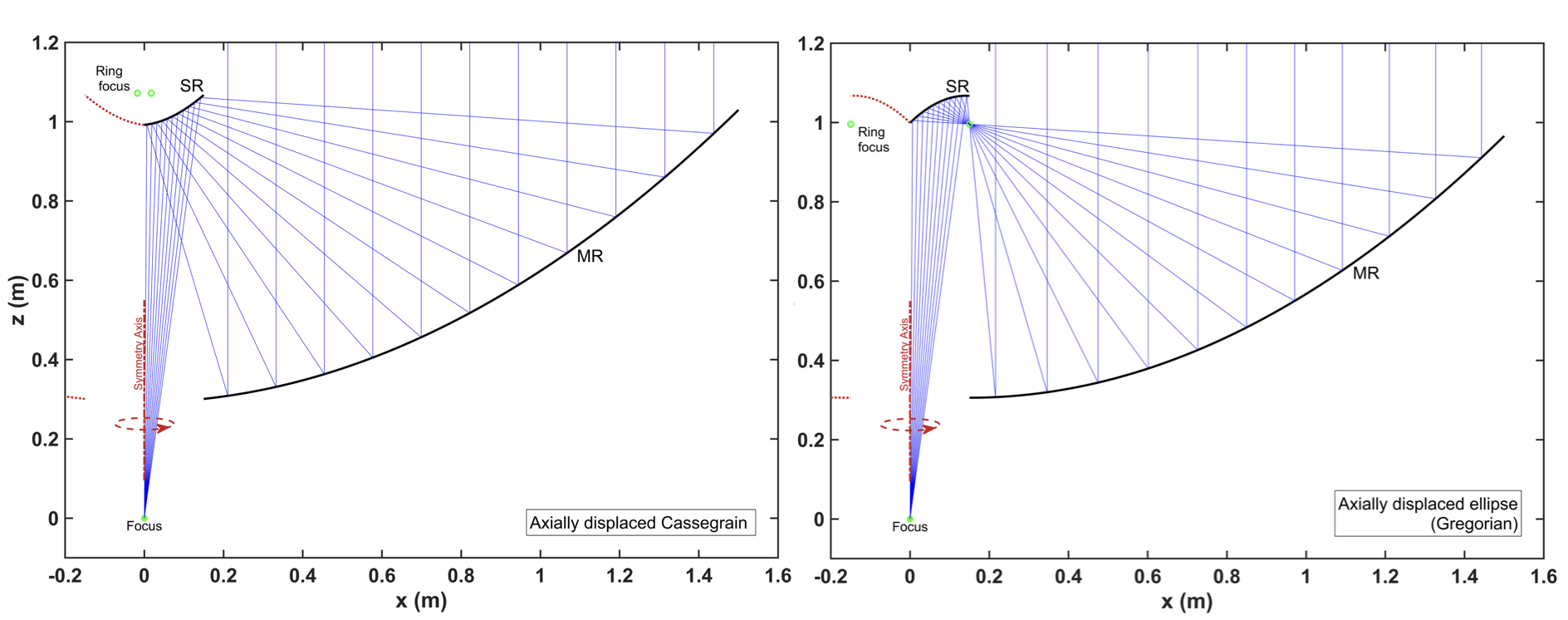}
   \end{tabular}
   \end{center}
   \caption[example] 
%>>>> use \label inside caption to get Fig. number with \ref{}
   { \label{fig:OpticalDesignConfigs} 
The Cassegrain and the axially displaced ellipse (Gregorian) optical configurations considered are shown.}
   \end{figure} 
   
Axially displaced ring focus designs are widely adopted in ground VLBI antenna and spacecraft communication HGA applications (e.g. VGOS, Europa Clipper, Roman Space Telescope) due to the high efficiencies they offer. These designs provide a clear optical path as opposed to the standard Cassegrain or Gregorian systems where reflections, due to the sub-reflector blockage, radiate back to the feed. This causes interference ripple over frequency in both the feed’s match and the radiation (gain). For these reasons, a ring focus dual reflector design was chosen for BHEX.

A parameter space exploration was carried out for Cassegrain and Gregorian configurations, shown in Fig.~\ref{fig:OpticalDesignConfigs}. The designs are defined by four geometric parameters \cite{GranetAntenna} - $D_m$  the diameter of the primary parabolic reflector, $F$, the focal distance of the primary reflector, $D_s$, the diameter of the subreflector (elliptical or hyperbolic) and $\theta_{e}$, the feed subtended half-angle. The full symmetrical reflector surface is generated by rotation about the axis offset from the focal axis of the primary reflector. In the Gregorian configuration, also called the axially displaced ellipse (ADE) design, the rays from the feed are inverted (Fig.~\ref{fig:OpticalDesignConfigs} right) leading to a more uniform aperture illumination resulting in high illumination efficiency compared to the Cassegrain system. This is the baseline design currently adopted for BHEX.

\begin{figure} [ht]
   \begin{center}
   \begin{tabular}{c} %% tabular useful for creating an array of images 
   \includegraphics[height=6.2cm]{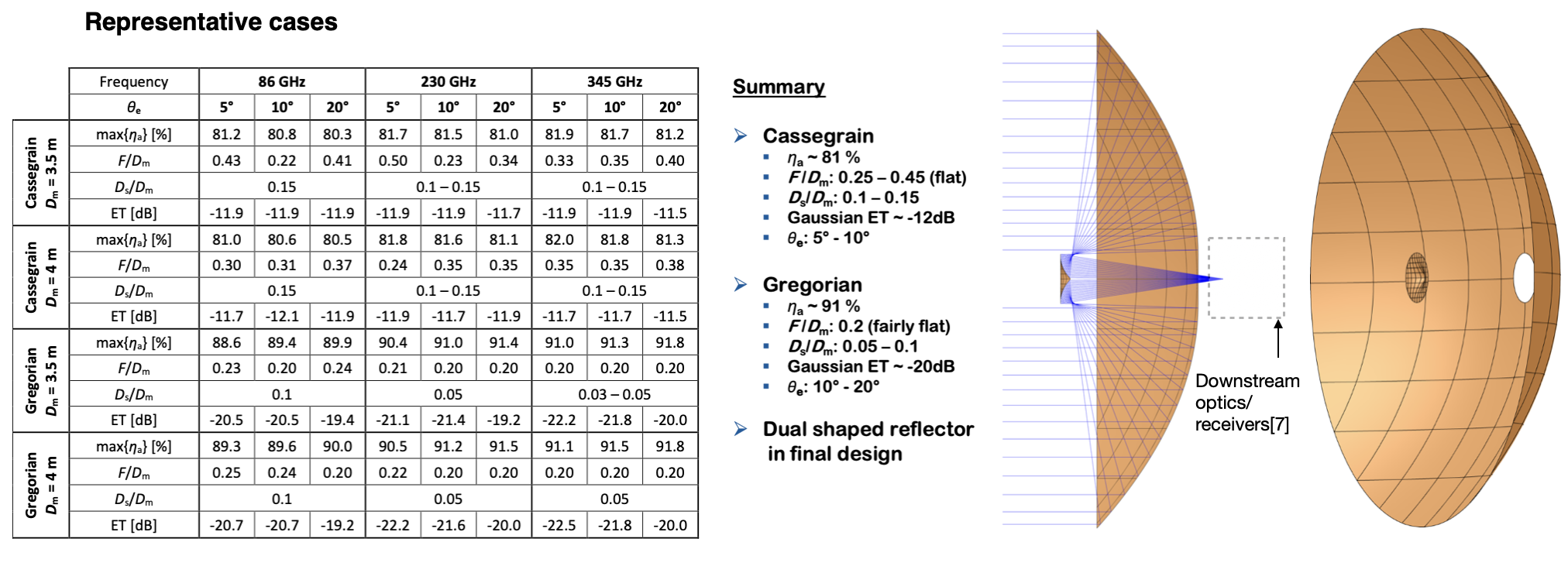}
   \end{tabular}
   \end{center}
   \caption[example] 
%>>>> use \label inside caption to get Fig. number with \ref{}
   { \label{fig:PrelimDesign} 
A summary of the preliminary optical design optimization results from a previous iteration for a 3.5/4 m aperture ({\it left}) and  a rendition of a current 3 m design ({\it right}) are shown. Further optimization and reflector shaping are in progress. The numbers listed assume an ideal reflector system, i.e. no surface deformations (Ruze losses) or anomalies due to subreflector support struts.}
\end{figure} 

A summary and representative cases from the parameter study are presented in Fig.~\ref{fig:PrelimDesign}. The parameter space explored was $F/D_m:0.2-0.5; ~ D_s/D_m:0.05-0.2; ~ \theta_e:5-20; ~ D_m: 3.5 ~\&~ 4$ m.  Design studies for a 3 m aperture are in progress and the results shown are from a previous design iteration. Fig.~\ref{fig:PrelimDesign} also shows a representative rendition of the current design that is not yet fully optimized. Further exploration of design choices and reflector shaping are currently being pursued, with results to be presented soon \cite{BHEX_IEEE_Lehmensiek_2024}.

\section{Future}

The planned next steps towards finalizing the design are the completion of the parameter optimization and shaping followed by structural and thermoelastic deformation studies under operational mission conditions, incorporating specific material properties, dimensions, the subreflector and its support structure. These studies will determine the allowed temperature distribution limits for maintaining the surface profile.  

\acknowledgements

This project/publication is funded in part by the Gordon and Betty Moore Foundation (Grant \#8273.01). It was also made possible through the support of a grant from the John Templeton Foundation (Grant \#62286).  The opinions expressed in this publication are those of the author(s) and do not necessarily reflect the views of these Foundations.
% References
\bibliography{report} % bibliography data in report.bib
%\bibliography{bhexspiepapers}
\bibliographystyle{spiebib} % makes bibtex use spiebib.bst

\end{document}